\documentclass[aps,pra,10pt,onecolumn,superscriptaddress]{revtex4}
\usepackage[paperwidth=220 mm,paperheight=297mm,centering]{geometry}
\usepackage{amsmath}
\usepackage{amssymb}
\usepackage{bbm}
\usepackage{graphicx}
\usepackage{framed}
\usepackage{color}
\usepackage{hyperref}

\newlength{\eqboxstorage}

\begin{document}

\title{Optimal processes for probabilistic work extraction beyond the second law}

\author{Vasco Cavina}
\affiliation{NEST, Scuola Normale Superiore and Istituto Nanoscienze-CNR, I-56126 Pisa, Italy}
\author{Andrea Mari}
\affiliation{NEST, Scuola Normale Superiore and Istituto Nanoscienze-CNR, I-56126 Pisa, Italy}

\author{Vittorio Giovannetti}
\affiliation{NEST, Scuola Normale Superiore and Istituto Nanoscienze-CNR, I-56126 Pisa, Italy}

\begin{abstract}
According to the second law of thermodynamics, for every transformation performed on a system which is in contact with an environment of fixed temperature, the extracted work is bounded by the decrease of the free energy of the system. However, in a single realization of a generic process, the extracted work is subject to statistical fluctuations which may allow for probabilistic violations of the previous bound. We are interested in enhancing this effect, {\it i.e.}\ we look for thermodynamic processes that maximize the probability of extracting work above a given arbitrary threshold. For any process obeying the Jarzynski identity, we determine an upper bound for the work extraction probability that depends also on the minimum amount of work that we are willing to extract in case of failure, or on the average work we wish to extract from the system. Then we show that this bound can be saturated within the thermodynamic formalism of quantum discrete processes composed by sequences of  unitary quenches and complete thermalizations.  We explicitly determine the optimal protocol which is given by two quasi-static isothermal transformations separated by a finite unitary quench. 
\end{abstract}
\maketitle

\noindent {\bf \large Introduction}\\

In classical thermodynamics \cite{huang1987} the (Helmholtz)   {\it free energy} of a system at thermal equilibrium is defined as $F:= U - T S$, where $U$ is the internal energy, $T$ is the temperature and $S$ is the entropy. Whenever the environment is characterized by a fixed and unique temperature $T$, in every process connecting two states having free energy $F_{\rm in}$ and $F_{\rm fin}$ respectively, the work done by the system is upper bounded by the free energy reduction
\begin{equation}
W  \le - \Delta F := F_{\rm in} - F_{\rm fin} . \label{2law}
\end{equation}
The previous inequality is a direct manifestation of the second law of thermodynamics. Indeed for a cyclic process $\Delta F=0$ and the bound \eqref{2law} states that no positive work can be extracted from a single heat bath. However, according to the microscopic theory of statistical mechanics \cite{huang1987}, the work done by a system in a given transformation is non-deterministic and can present statistical fluctuations \cite{ bochkov1977, esposito2009,evans1993,wang2002}. For macroscopic systems, like a gas in contact with a moving piston, these fluctuations are usually negligible and one can replace all random variables with their averages recovering the thermodynamic bound \eqref{2law}. For sufficiently ``small" systems, {\it i.e.} those in which the number of degrees of freedom are limited and the energies involved are of the order of $k_B T$, fluctuations are important and for each repetition of a given protocol the system can produce a different amount of work, which can be described as a random variable with an associated probability distribution \cite{evans1993,wang2002,esposito2009}. Moreover a further source of difficulty in the description of ``small" systems, like nano-scale devices, molecules, atoms, electrons, {\it etc.}, is that quantum effects are often non-negligible and  quantum fluctuations can also affect the work extraction process \cite{horodecki2013,vinjanampathy2015,goold2015, skrzypczyk2014}. 

The properties and the constraints characterizing the work probability distribution of a given process are captured by the so called {\it fluctuation theorems} which can be defined both for classical \cite{bochkov1977, esposito2009,crooks2000, jarzynski1997,jarzynski2011} and quantum systems \cite{tasaki2000, campisi2011, esposito2006,aberg2016}. These theorems can be seen as generalizations of the second law  for  processes characterized by large statistical fluctuations. Indeed it can be shown \cite{jarzynski1997,jarzynski2011} that the expectation value of the work distribution always satisfies the bound \eqref{2law}, while for a single-shot realization of the protocol it is in principle possible to extract a larger amount of work at the price of succeeding with a small probability \cite{evans1993,wang2002, jarzynski2011, alhambra2015, halpern2015}.  

Our aim  is to identify the optimal protocols for maximizing the probability of extracting work above a given threshold $\Lambda$, arbitrarily larger than the bound \eqref{2law}. Differently from standard thermodynamics in which the optimal procedures are usually identified with the quasi-static (reversible) transformations saturating the inequality \eqref{2law}, for the problem we are considering, fluctuations are necessary in order to probabilistically violate the second law. Consequently, in this case optimality will require some degree of irreversibility. 
 In our analysis we shall focus on processes obeying the Jarzynski identity~\cite{jarzynski1997,tasaki2000,esposito2006,campisi2011} which include all those transformations where a system originally at thermal equilibrium evolves under an externally controlled, time-dependent Hamiltonian and proper concatenations of similar transformations.  In this context, as a first step we identify an upper bound for the probability of work extraction above the threshold $\Lambda$ which depends on the minimum amount of work $W_{\rm min}$  that we are willing to extract in case of failure of the procedure. 
 We also identify the class of optimal protocols that  enable one to saturate such bound. 
 These  correspond to have two quasi-static transformations separated by a single, abrupt modification of the Hamiltonian (unitary quench),  the associated work distribution being characterised by only two possible outcomes: one arbitrarily above the bound~\eqref{2law} (success) and one below (failure).  Explicit examples are presented in the context  of discrete thermal processes~\cite{aberg2013,anders2013} and in the context of  one-molecule Szilard-like heat engines~\cite{SZILARD}.

In the second part of the paper we focus instead on the upper bound for the probability of work extraction 
above a given threshold $\Lambda$ which applies to all those processes that ensure a fixed value $\mu$ of the 
average  extracted work. Also in this case we present explicit protocols which enable one to saturate the bound: it turns
out that they belong to the same class of the optimal protocols we presented in the first part of the manuscript (i.e. 
 two quasi-static transformations separated by a single, unitary quench). \\

%The paper is organized as follows. We start in Sec.~\ref{sec:BOUND}   by reviewing some basic facts about the Jarzynski identity and by deriving the bound on the work
%extraction probability under the constraint on the minimum amount of work. 
% In Sec.~\ref{sec:DISC} we then present two different models which allows us to identifies the optimal processes which 
% saturate the bound. In Sec.~\ref{sec:AVE} we then study the optimal probability distribution for extracting work above a certain
% threshold under the constraint on the average work.
% Conclusions and perspectives are given in Sec.~\ref{sec:CON} while technical aspects are presented in the Appendix. 

\vspace{1 em}
\noindent {\bf \large Work extraction above threshold under minimal work constraint}\\
%\label{sec:BOUND}

In 1997 \cite{jarzynski1997}, motivated by molecular biology experiments,  C. Jarzynski derived an identity characterizing the probability distribution  $P(W)$ of the work  done by a system which is initially in a thermal state of temperature $T$ and it is subject to an Hamiltonian process interpolating from $H_{\rm in}$ to $H_{\rm fin}$. The identity is the following:
\begin{equation}
e^{- \beta \Delta F} =  \langle e^{\beta W} \rangle:= \int_{-\infty}^{\infty} P(W) e^{\beta W} dW, 
 \label{jarzynski}
\end{equation}
where  $\beta=\tfrac{1}{k_B T}$, $\Delta F=F_{\rm fin} -F_{\rm in}$,  with $F_{\rm in}, F_{\rm fin}$ being the free energies associated to the canonical thermal states with Hamiltonian $H_{\rm in}$ and $H_{\rm fin}$ respectively.
Notice that, while $F_{\rm in}$ corresponds to the actual free energy of the initial state, $F_{\rm fin}$ is not directly related to the final state  which may be out of equilibrium but only to its final Hamiltonian (in all the processes considered in this work, however  the initial and final states are always thermal, and $F_{\rm fin}$ is also the actual free energy of the system). Notice also that by a simple convexity argument applied to the exponent 
on the right-hand-side of  Eq.~(\ref{jarzynski}) one gets the inequality 
\begin{equation}
 \langle W \rangle:= \int_{-\infty}^{\infty} P(W) W  dW \leq -  \Delta F, 
 \label{SECOND}
\end{equation}
which is the counterpart of (\ref{2law}) for non-deterministic  processes.

For quantum systems, because of the intrinsic uncertainties characterizing quantum states, identifying a proper definition of work is still a matter of research \cite{horodecki2013, skrzypczyk2014, gallego2015, roncaglia2014, talkner2007}. However, if we assume to perform a measurement of the energy of the system before and after a given unitary process, the work extracted during the process is operationally well defined as the decrease of the measured energy. Also in this quantum scenario, it can be shown \cite{tasaki2000,esposito2006,campisi2011} that the work distribution obeys the Jarzynski identity \eqref{jarzynski}.

Among all (unitary or non-unitary) processes interpolating from an initial Hamiltonian $H_{\rm in}$ to a final Hamiltonian $H_{\rm fin}$ and fulfilling~(\ref{jarzynski}), we are interested in maximizing the probability of extracting work above a given arbitrary threshold $\varLambda$, i.e.
in computing the quantity 
\begin{equation}
P_{\max} (W \ge \varLambda) :=  \max_{\rm all\ processes}  P(W \ge \varLambda), \label{Pmax}
\end{equation} 
where 
\begin{equation}
P(W \ge \varLambda) := \int_{\varLambda}^{+\infty} P(W) d W. \label{PLambda}
\end{equation} 

This problem is particularly interesting and non-trivial only when the threshold is beyond the limit \eqref{2law} imposed by the second law, {\it i.e.}\  when $\varLambda > -\Delta F$. Indeed for lower values of $\varLambda$, it is known that the probability \eqref{Pmax} can be maximized to $1$ by an arbitrary thermodynamically reversible (quasi-static) process, deterministically extracting $W=-\Delta F$. Exploiting statistical fluctuations, larger values of work can be extracted \cite{evans1993,wang2002, jarzynski2011, alhambra2015, halpern2015}. But what are the corresponding probabilities and the associated optimal processes?

To tackle this problem we start considering those processes which, beside fulfilling~(\ref{jarzynski}), satisfy also the constraint 
\begin{eqnarray} 
P(W)=0\;,  \qquad \forall  \;W< W_{\rm min}\;, \label{constr1}
\end{eqnarray} 
with $W_{\rm min}$ an assigned value smaller than or equal to $\Lambda$ (consistency condition being $\Lambda$ 
the work threshold above which we would like to operate) and
 $-\Delta F$ (by construction, being the latter an upper bound to the average work of the process, see Eq.~(\ref{SECOND})). This corresponds to set a lower limit on the
 minimum amount of work that we are willing to extract in the worst case scenario, the unconstrained scenario being recovered
 by setting $W_{\rm min} \rightarrow -\infty$ (instead taking $W_{\rm min}=0$ we select those processes where, in all the statistical realizations   no work is ever provided to the system).
 
Under the above hypothesis  the following inequality can be established 
\begin{equation} 
 P(W \geq \varLambda) \leq \frac{ e^{- \beta \Delta F} - e^{\beta W_{\rm min}}}{e^{\beta \varLambda} -e^{\beta W_{\rm min}}},\label{bound}
  \end{equation}
 which
  in the unrestricted regime 
  $W_{\rm min} \rightarrow -\infty$, yields 
\begin{equation}  
 P(W \geq \varLambda ) \leq e^{-\beta (\varLambda + \Delta F)},  \label{bound2}
 \end{equation}
 that was already demonstrated in \cite{jarzynski2011}.
 Again we stress that both bounds \eqref{bound} and \eqref{bound2} are relevant only for $\Lambda > -\Delta F$, while for $\Lambda \le -\Delta F$ they  can be trivially replaced by $P(W \geq \varLambda) \leq 1$.
 
\begin{figure}[t]
\includegraphics[width=0.44  \textwidth]{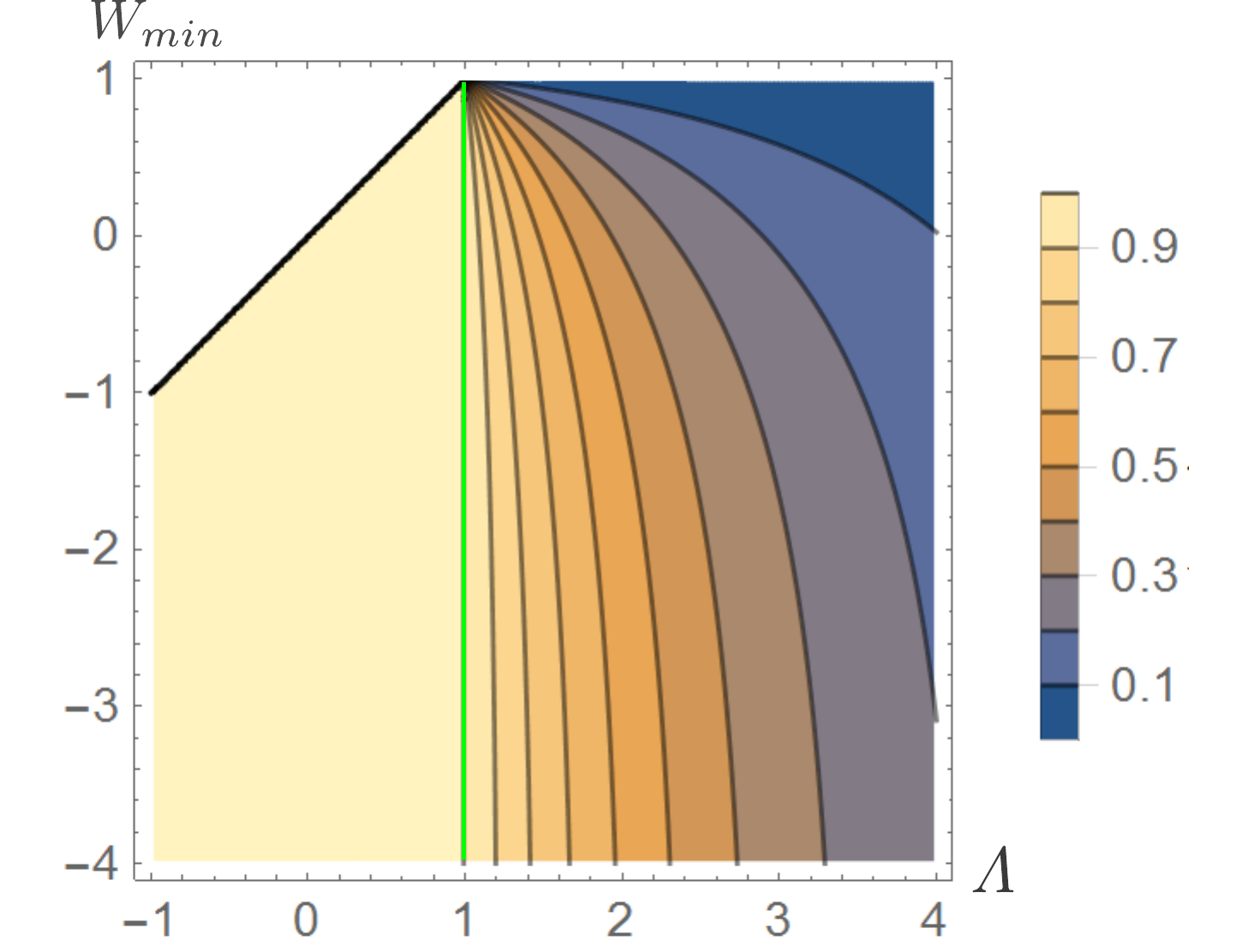}
\caption{Contourplot of the optimal probability $P_{\max}^{(W_{\rm min})}(W \ge \varLambda)$ of Eq.\ \eqref{Pmax11} as a function
of $\Lambda$ and $W_{\rm min}$ for $\beta =\frac{1}{2}$ and $-\Delta F=1$. 
If $\Lambda \leq -\Delta F$ the probability of success is equal to $1$ independently from $W_{\rm min}$.
For $\Lambda \geq -\Delta F$ the bound is non trivial and is given by Eq. (\ref{bound}).
Values of $W_{\rm min}$ 
larger than $\Lambda$ or larger than $-\Delta F$ needs not to be considered. }
\label{contours1}
\end{figure}

The proof of the general bound \eqref{bound} follows straightforwardly from the identity \eqref{jarzynski} and the definitions of $\varLambda$ and $W_{\rm min}$.
 Indeed, we can split the integral appearing in \eqref{jarzynski} as the sum of  two terms that can be independently bounded as follows: 
 \begin{eqnarray}  
 e^{-\beta \Delta F}& =& \int_{ W_{\min}}^{\varLambda}\!\!\! \!\!\!e^{\beta W}   P(W) dW  + \int_{\varLambda}^{\infty} \!\!\!\!\!e^{\beta W}  P(W) dW \nonumber \\
& \geq & e^{\beta W_{\rm min}} \int_{ W_{\min}}^{\varLambda}  \!\!\! \!\!\!  P(W) dW  + e^{\beta \varLambda}  \int_{\varLambda}^{\infty}  \!\!\! \!\!\!P(W) dW \nonumber \\\label{step2}
& = &  e^{\beta W_{\rm min}} (1-P(W \ge \varLambda))  + e^{\beta \varLambda} P(W \ge \varLambda
 ),
 \end{eqnarray}
 Equation~\eqref{bound} finally follows by solving the resulting inequality for $P(W \ge \varLambda)$.
 We summarize the bounds given in this this paragraph in Fig. (\ref{contours1}).\\
 
\vspace{1 em}
\noindent {\bf  Attainability of the bound~(\ref{bound})}\\
%\label{sec:DISC}
 By close inspection of the derivation~(\ref{step2}) it is clear that the only way to saturate the inequality~(\ref{bound}) is by means 
 of processes whose work distribution is the convex combination of two delta functions centered respectively at 
$W=W_{\rm min}$ and  $W=\Lambda$,  i.e. 
 \begin{equation}
P(W)= p  \delta  (W- \varLambda)+(1-p) \delta  (W- W_{\rm min}), \label{PWtot2}
\end{equation}
with $p=P(W \ge \varLambda)$ equal to the term on the rhs of Eq.~(\ref{bound}), i.e. 
 \begin{equation}
p=  \frac{ e^{- \beta \Delta F} - e^{\beta W_{\rm min}}}{e^{\beta \varLambda} -e^{\beta W_{\rm min}}} \label{poptimal}.
\end{equation}
Yet it is not obvious  whether such transformations exist nor how one could implement them while still obeying the Jarzynski identity.
To deal with this question in the following we present two different schemes both capable of fulfilling these requirements
proving hence that
the following identity holds:
\begin{equation} 
P_{\max}^{(W_{\rm min})}(W \ge \varLambda)  = \frac{ e^{- \beta \Delta F} - e^{\beta W_{\min}}}{e^{\beta \varLambda} -e^{\beta W_{\min}}}, \label{Pmax11}
\end{equation} 
(see Fig. \ref{contours1} for a contour plot of this optimal probability).
The first example is based on a specific theoretical framework (the discrete quantum process approach) introduced in Ref.~\cite{anders2013}
for modeling thermodynamic transformations applied to quantum systems. While the attainability of (\ref{bound}) does not require to consider a full
quantum treatment (only the presence of discrete energy exchanges being needed, not quantum coherence), the use of this technique turns out to be useful as it provides a simplified, yet fully exhaustive description of the involved transformations. The second example is instead fully classical and it is based on an idealized
one-molecule Szilard-like heat engine.

\vspace{1 em}
\noindent {\bf  Optimal transformations by discrete quantum processes}\\
Following the approach presented in Refs.~\cite{aberg2013,anders2013} in this section we consider protocols composed by the concatenation of only two types of operations:  {\it discrete unitary quenches} (DUQs) and {\it discrete thermalizing transformations} (DTTs). 
 A DUQ applied to a system described by an input density matrix $
\rho$ and Hamiltonian $H$,  is an arbitrary change of the latter  which does not affect the former, i.e. 
\begin{eqnarray}
&& H \xrightarrow{DUQ} H'\;,  
\nonumber \\
&& \rho \xrightarrow{DUQ}  \rho' =
\rho \;,
\end{eqnarray}
$H'$ being the final Hamiltonian  of the system. 
A DTT instead is a complete thermalization towards a Gibbs state of temperature $T$ without changing the system Hamiltonian, i.e. 
\begin{eqnarray}
 H  &\xrightarrow{DTT}& H' = H  \;,  \nonumber 
\\
\rho &\xrightarrow{DTT}&
\rho' =  \omega^{(\beta)}_{H}\;,
\end{eqnarray}
with  $\omega^{(\beta)}_{H} :={e^{-\beta H}}/{Z}$, 
where  $\beta:=\tfrac{1}{k_B T}$ and  $Z:={\rm Tr}[ e^{-\beta H}]$  
the associated  partition function.
Operationally, a DUQ can be implemented by an instantaneous change of the Hamiltonian realized while keeping the system thermally isolated. A DTT instead can be obtained by weakly coupling the system with the environment for a sufficiently long time. The convenience for introducing such elementary processes is that the energy exchanged during a DUQ and a DTT can be thermodynamically interpreted as work and heat respectively, without any risk of ambiguity
typical of continuous transformations in which the Hamiltonian and the state are simultaneously changed. 
On the other hand, as shown in \cite{aberg2013,anders2013}, continuous transformations can be well approximated by a sequence of
infinitesimal DUQs and DTTs.

In order to show that Eq.~(\ref{bound}) can be saturated we then focus on a $N$-long sequence of alternating DUQs and DTTs operated at the same temperature $T$ and connecting
an input  Gibbs state  $\omega^{(\beta)}_{H_{\rm in}}$
to a final Gibbs state $\omega^{(\beta)}_{H_{\rm fin}}$ via the following steps  
\begin{equation}
\begin{array}{ccc}
{\rm Step}\ 1: &  H_0 \xrightarrow{DUQ} H_1,  & \omega^{(\beta)}_{H_0} \xrightarrow{DTT} \omega^{(\beta)}_{H_1}, \\
{\rm Step}\ 2 : &H_1 \xrightarrow{DUQ} H_2 ,  & \omega^{(\beta)}_{H_1} \xrightarrow{DTT} \omega^{(\beta)}_{H_2}, \\ 
\vdots \\
{\rm Step}\ j:  & H_{j-1} \xrightarrow{DUQ} H_j,  &  \omega^{(\beta)}_{H_{j-1}} \xrightarrow{DTT} \omega^{(\beta)}_{H_j}, \\
\vdots \\
{\rm Step}\ N:  & H_{N-1} \xrightarrow{DUQ} H_N,  & \omega^{(\beta)}_{H_{N-1}} \xrightarrow{DTT} \omega^{(\beta)}_{H_N},
\end{array} \label{steps}
\end{equation}
where for $j=0, 1,\cdots, N$,  $H_j$ represents the Hamiltonian of the system at the end of the DUQ of the $j$-th step and where for easy of notation we set 
$H_0:= H_{\rm in}$ and $H_N: = H_{\rm fin}$.  
As further assumption we shall also restrict the analysis to those cases where all the $H_j$ entering the sequence (the initial and the final one included) mutually commute, i.e. 
$[H_j,H_{j+1}]=0$. Accordingly 
 the action of each DUQ corresponds to a simple shift of the energy levels without changing the corresponding eigenstates:
\begin{equation}
H_j=\sum_k E_k^{(j)}  |k\rangle \langle k|   
\quad  \xrightarrow{DUQ} \quad H_{j+1}=\sum_k E_k^{(j+1)}   |k\rangle \langle k|.
\end{equation}
Also, since all the  Gibbs states  $\omega^{(\beta)}_{H_j}$ are diagonal in the same energy basis $\{  |k\rangle\}$, 
quantum coherence will not play any role in the process,  meaning that the results we obtain could be directly applicable to classical models.
  This is not a strong limitation since, as we are going to show, thermodynamically optimal processes  saturating~(\ref{bound}) are already obtainable within this limited set of semi-classical operations.

To determine the probability distribution of work for a generic sequence~(\ref{steps}) 
observe that in the $j$-th step work can be extracted from (or injected to) the system only during the associated DUQ~\cite{anders2013}. Here the state  is described by the density matrix 
$\omega^{(\beta)}_{H_{j-1}}=\sum_k p_k^{(j-1)}  |k\rangle \langle k|$, 
with \begin{eqnarray} p_k^{(j-1)} &:=& e^{- \beta E_k^{(j-1)}} / Z_{j-1}, \label{DEFP}
\end{eqnarray}  being the probability of finding it into the $k$-th energy eigenstate whose energy 
passes from $E_k^{(j-1)}$ to $E_k^{(j)}$ during 
the quench. When this happens the system acquires a $\Delta E_k^{(j)}:= E_k^{(j)} -E_k^{(j-1)}$ increment of internal energy, corresponding to 
an amount of $- \Delta E_k^{(j)}= E_k^{(j-1)} -E_k^{(j)}$ of work production (the system being thermally isolated during each DUQ). 
Accordingly the probability distribution of the work done by the system during the $j$-th step  can be expressed as
 \begin{equation} 
 P_j(W_j) = \sum_k   p_k^{(j-1)}    \delta(W_j +\Delta E_k^{(j)}). \label{PDUQ}
 \end{equation}
At the next step the  system first thermalizes via a DTT which, independently from the previous history of the process, brings in the Gibbs state $\omega^{(\beta)}_{H_{j}}$,
and then undergoes to a new DUQ that produces an extra amount of  work $W_{j+1}$ whose statistical distribution $P_{j+1}(W_{j+1})$ can be expressed 
as in  Eq.~(\ref{PDUQ}) by replacing $j$ with $j+1$. 
The total work $W$ extracted during the whole transformation~(\ref{steps})   can finally be computed by summing all the $W_j$'s, 
the resulting probability distribution being
  \begin{eqnarray} 
 P(W) &=& \int d W_1 \cdots \int  dW_N 
 P_1(W_1) \cdots P_N(W_N)   \delta(W- (W_1 + \cdots + W_N)) \label{PDUQ1},
 \end{eqnarray}
 which can be easily shown to satisfy the Jarzynski identity \eqref{jarzynski} (see section Methods) and hence the inequality~(\ref{bound})
 which follows from it. 
 \begin{figure}[t]
\includegraphics[width=0.44 \textwidth]{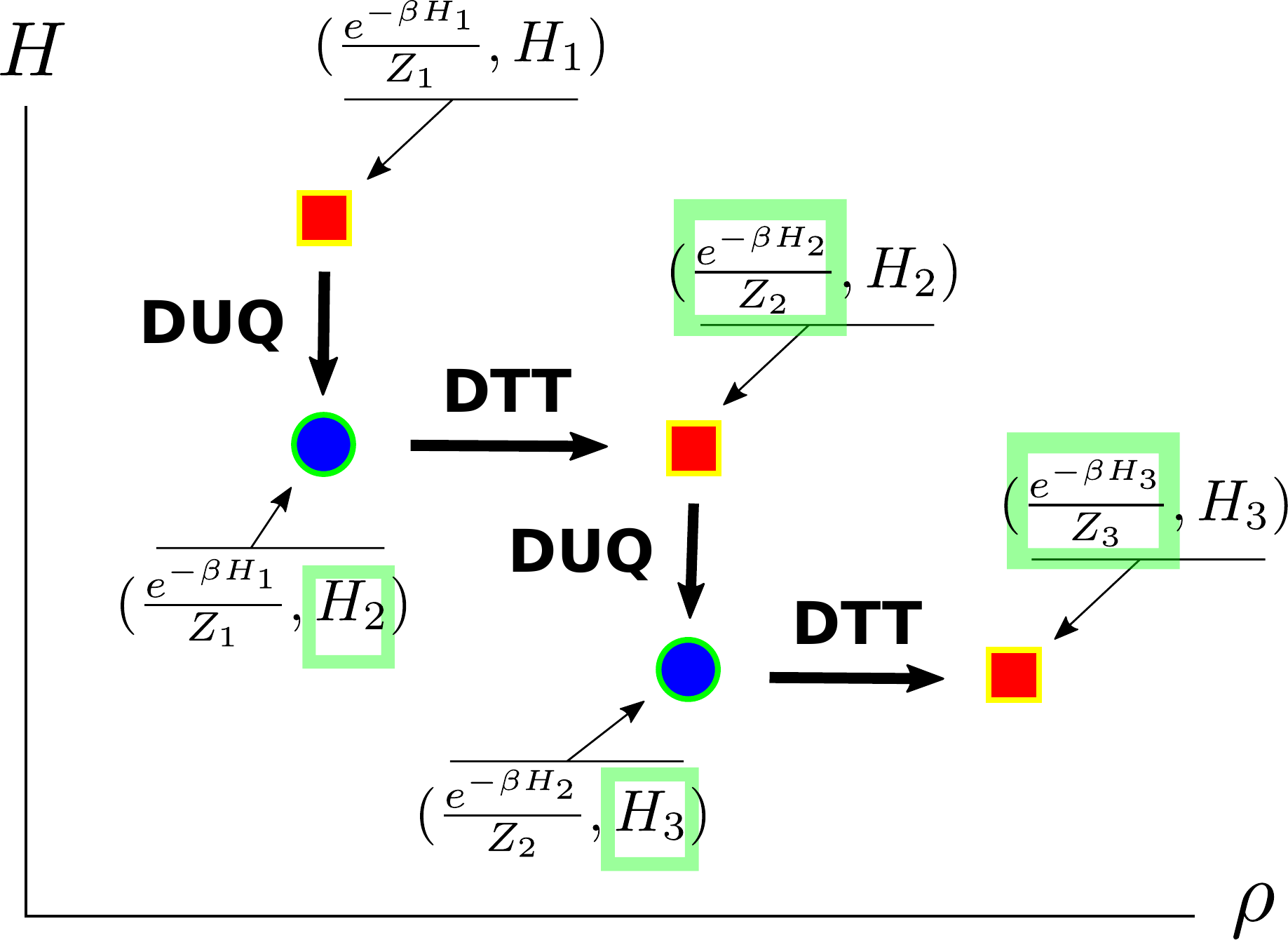}
\caption{Sequence of two ``steps", each one composed of a DUQ and a DTT. Each configuration point $(\rho,H)$ is represented by a red square if it is
an equilibrium configuration $(\omega_H,H)$, and by a blue circle otherwise.}\label{fig:staircase}
\end{figure}

It is a basic result of thermodynamics and statistical mechanics that the inequality \eqref{2law} can be saturated by isothermal transformations in which the system is changed very slowly in such a way that its state remains always in equilibrium with the bath \cite{huang1987}. These operations are usually called {\it quasi-static} or {\it reversible}. In the  framework of discrete quantum processes quasi-static transformations can be obtained in the limit of infinite steps $N\rightarrow \infty$ while keeping fixed the initial  
and final   Hamiltonian of the sequence. Indeed, as shown in \cite{aberg2013,anders2013}, interpolating between the initial and final Hamiltonian by a sequence of infinitesimal changes (e.g. fulfilling the constraint $\beta \sum_k |\Delta E_{k}^{(j)}| \ll 1$) each followed by complete thermalizations,
 one can saturate the bound \eqref{2law}. 
In terms of the probability distribution~(\ref{PDUQ1}) one can easily show \cite{aberg2013} that in the quasi-static limit we obtain a delta function centered in $-\Delta F = F_{\rm in} - F_{\rm fin}$, i.e. 
\begin{equation}
P(W)\Big|_{N\rightarrow \infty} =\delta(W+ \Delta F), \label{Pdelta}
\end{equation}
which means that for every realization of the process the work extracted is the maximum allowed by the second law \eqref{2law} with negligible fluctuations. This can be understood from the fact that the total work $W$ is the sum of $N$ independent random variables $W_j$ and therefore we expect the fluctuations around the mean $\langle W \rangle$ to decay as $1/\sqrt{N}$. Moreover, since the Jarzynski identity \eqref{jarzynski} must hold, the only possible value for the mean of an infinitely sharp  distribution is $\langle W \rangle=-\Delta F$.

We can now come back to our original problem of determining the maximum probability of extracting an amount of work larger than an arbitrary value $\varLambda$, for fixed values of the initial and final Hamiltonians $H_{\rm in}$ and $H_{\rm fin}$. 

If the threshold is below the free energy decrease of the system, {\it i.e.} if $\varLambda \le -\Delta F$, the problem is trivial. 
In this case a quasi-static transformation interpolating between the initial and final Hamiltonian is optimal. Indeed, as expressed in Eq.\ \eqref{Pdelta}, the work extracted in the process is deterministically equal to $- \Delta F$ which is larger than $\Lambda$.
Formally, integrating \eqref{Pdelta}, we have that for a quasi-static process
 
\begin{equation} 
P ( W \geq \varLambda)  = \Bigg\{
\begin{array}{ll}
 1, &  {\rm for } \  \varLambda \le -\Delta F, \\ 
 & \\
 0, & {\rm for} \ \varLambda  > -\Delta F.
\end{array}   \label{Ptrivial}
\end{equation} 
The cumulative work extraction probability \eqref{Ptrivial} shows that, despite quasi-static processes are optimal for $\Lambda \le -\Delta F$, they are absolutely useless for $\Lambda > -\Delta F$ where the probability drops to zero. Then, if we want to explore the region $\varLambda > -\Delta F$ which is beyond the limit imposed by the second law, it is clear that we have to exploit statistical fluctuations typical of non-equilibrium processes. 

%%%
\begin{figure}[t]
\includegraphics[width=0.44 \textwidth]{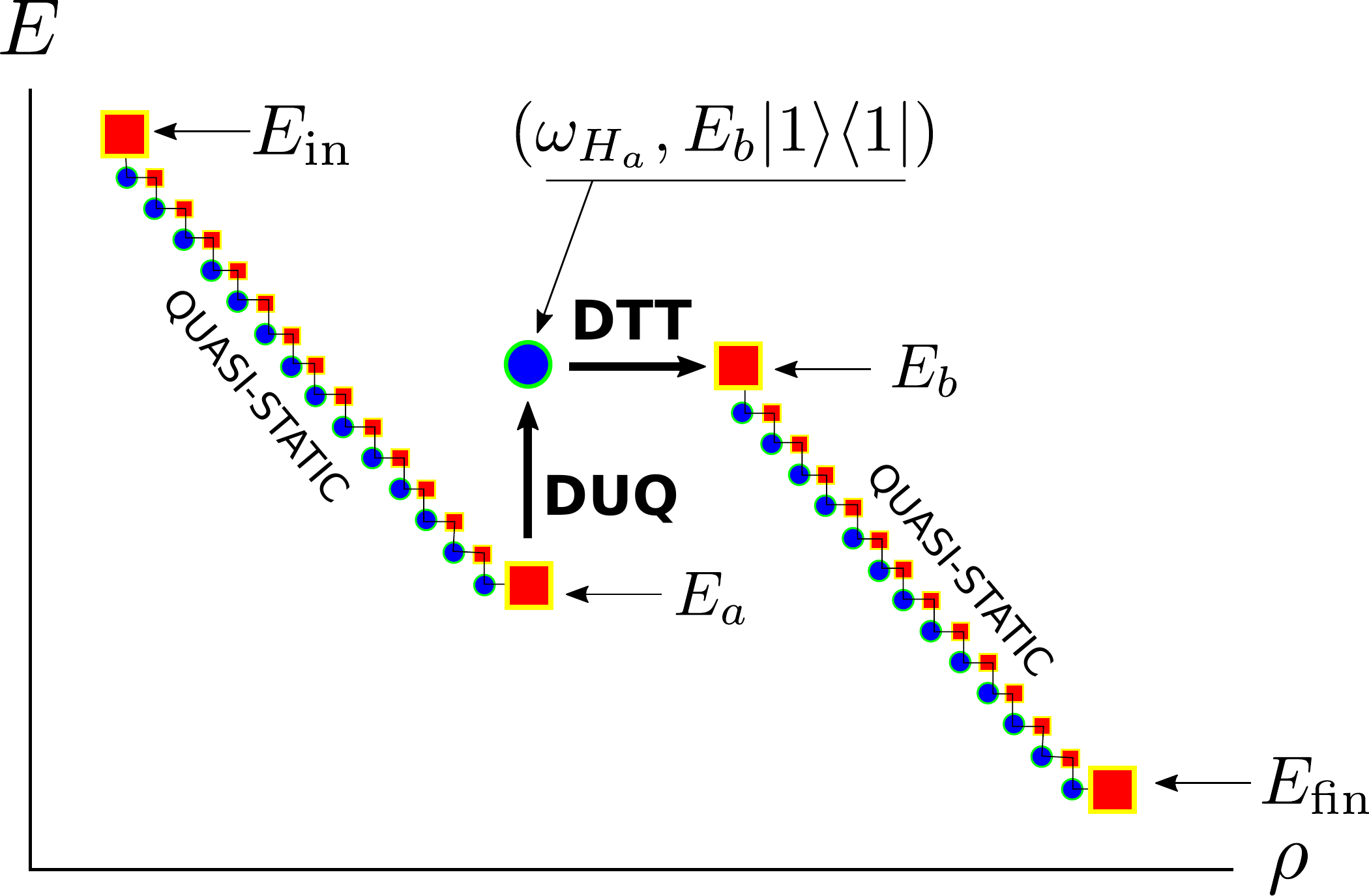}
\caption{
Scheme of the optimal process saturating the bound \eqref{bound}, valid for a two-level system with Hamiltonian $H=E_1 |1\rangle \langle 1|$.  The process is divided in three steps: a) a
quasi-static transformation where the value of $E_1$ varies from  $E_{\rm in}$ to $E_a$, b) an Hamiltonian quench from  $E_a$ to $E_b$ followed by a thermalization, c) a final quasi-static transformation from $E_b$ to $E_{\rm fin}$. }
\label{fig:4}
\end{figure}
%%%%%%%%

Consider next the case of a threshold larger than the decrease of free energy $\varLambda > -\Delta F$.
To identify an optimal process fulfilling~(\ref{bound}) it is sufficient to focus on the simplest scenario of a two-level system with energy eigenstates $|0\rangle$ and $|1\rangle$ and such that the eigenvalue associated with $|0\rangle$ always nullifies, i.e.  $E_0^{(j)}=0$ for all $j$.
Accordingly each Hamiltonian and corresponding Gibbs state can be expressed in 
terms of a single real parameter $E_1^{(j)}$:  
\begin{equation}
H_j= E_1^{(j)} |1\rangle \langle 1|, \quad \omega^{(\beta)}_{H_j}= \frac{  |0 \rangle  \langle 0| + e^{-\beta E_1^{(j)} } |1 \rangle  \langle 1|}{1+e^{-\beta E_1^{(j)} } }.
\end{equation} 
In this way the generic process described in (\ref{steps}) is completely characterized by assigning a sequence of $N+1$ parameters $\{ E_1^{(0)} , E_1^{(1)} ,E_1^{(2)}, \cdots,  E_1^{(N)} \}$ arbitrarily interpolating between the initial value  $E_1^{(0)} =E_{\rm in}$ and the final value $E_1^{(N)} = E_{\rm fin}$.

Let us then focus on the protocol composed by the following three steps and summarized in  Fig.~\ref{fig:4}: 
\begin{enumerate}
\item[a)] perform a quasi-static transformation from the initial value $E_{\rm in}$ to the value $E_a$ to be fixed later on:
\begin{equation}
\{ E_{\rm in}, E_{\rm in}+\epsilon , E_{\rm in}+2 \epsilon, \dots, E_a \}, \label{sequence1} 
\end{equation}
with $\epsilon$ being a small increment which we shall send to zero while sending the associated number of steps $N_a = 
|E_a - E_{\rm in}|/ |\epsilon|$  to infinity;
\item[b)] apply a finite DUQ  from $E_a$ to another arbitrary value $E_b> E_a$ also to be fixed later on, 
followed by a complete thermalization:
\begin{equation}
E_a \xrightarrow{DUQ+DTT} E_b\;;
\end{equation}
\item[c)] perform a quasi-static transformation from $E_b$ to the desired final value $E_{\rm fin}$:
\begin{equation}
\{E_b, E_b+ \epsilon, E_b+2  \epsilon, \dots, E_{\rm fin} \},
\end{equation}
where, as in step a), $\epsilon$ is a small increment which we shall send to zero by sending 
the associated number of steps $N_c = |E_{\rm fin} - E_b|/ |\epsilon|$  to infinity.
\end{enumerate}
In the limit  $\epsilon \rightarrow 0$, since the initial and final configurations are fixed the only free parameters of this protocol are $E_a$ and $E_b$, and they will affect the final probability distribution of the work
done by the system. 
In particular according to Eq.~(\ref{Pdelta}), the work extracted in the two quasi-static transformations a) and c) is deterministically given by the corresponding free energy reductions, i.e. the quantities 
\begin{eqnarray} 
W_a &=& F_{\rm in}-F(E_a), \nonumber \\
W_c &=&  F(E_b)-F_{\rm fin},\label{red}
\end{eqnarray} 
 respectively, with $F(E):= - \frac{1}{\beta}\log(e^{-\beta E} +1)$.   The work extracted in the intermediate operation b) instead  is $W_b=0$ if the system is in the the state $|0\rangle$ (this happens with probability $p=\frac{1}{1+e^{-\beta E_a}}$) while it is equal to the negative quantity $W_b=E_a-E_b$ if the system is in the state $|1\rangle$ (which happens with probability 
$1-p$). Accordingly the total work $W=W_a+W_b+W_c$ 
 is a convex combination of two delta functions:
\begin{equation}
P(W)=p \delta  (W- W_{\rm max})+(1- p)\delta  (W- W_{\rm min}), \label{PWtot}
\end{equation}
where 
\begin{eqnarray}
W_{\rm max} &:=& -  \Delta F  +F(E_b) -F(E_a) > - \Delta F,  \qquad \label{eqLambda}\\
W_{\rm min} &:=& W_{\rm max} + E_a - E_b < - \Delta F . \label{eqWmin}
\end{eqnarray}
Equation~(\ref{PWtot})  is of the form required to saturate~(\ref{bound}), see Eq.~(\ref{PWtot2}). 
 Indeed from  Eqs.~\eqref{eqLambda} and \eqref{eqWmin} it is easy to check that all values of $W_{\rm max} > - \Delta F$ and $W_{\rm min} < -\Delta F$ can be obtained by properly choosing $E_a$ and $E_b$, with $E_b > E_a$.
Moreover the function that gives $W_{\min}$ and $W_{\max}$ from $E_a$ and $E_b$ is a bijection, that can be inverted obtaining:
\begin{equation} \label{eEa}   e^{-\beta E_a} = \frac{e^{-\beta \Delta F} - e^{\beta W_{\max}}}{e^{\beta W_{\min}} - e^{-\beta \Delta F}},      \end{equation}
\begin{equation} \label{eEb}   e^{-\beta E_b} = \frac{e^{\beta \Delta F} - e^{-\beta W_{\max}}}{e^{-\beta W_{\min}} - e^{\beta \Delta F}},         \end{equation}
and hence
\begin{equation} \label{PPPP} 
p=\frac{ e^{- \beta \Delta F} - e^{\beta W_{\min}}}{e^{\beta W_{\max}} -e^{\beta W_{\min}}},
\end{equation}
the positivity of this expression being guaranteed by the ordering 
\begin{eqnarray} W_{\min} \leq -\Delta F \leq W_{\max}\label{ordering}, \end{eqnarray}  
which naturally follows from Eq.~(\ref{SECOND}). 
Equations~(\ref{PWtot2}) and (\ref{poptimal}) are finally obtained from~(\ref{PWtot}) and (\ref{PPPP}) 
by simply taking $W_{\rm max}=\varLambda$.

 From the above analysis it is evident that optimal processes saturating~(\ref{bound}) 
 can be obtained only for transformations 
that are quasi-static apart from a single finite DUQ which introduces a single probabilistic dichotomy on the final work distribution as required
by Eq.~(\ref{PWtot2}).
For a two-level system, we have just shown that they are characterized by 
the two parameters $E_a$  and $E_b$ (the values of $E_{\rm in}$ and $E_{\rm fin}$  being fixed by the initial and 
final Hamiltonians). However their choice is also completely determined by the desired maximum and minimum work values entering~(\ref{bound}): $\varLambda$ and $W_{\rm min}$.
Therefore we conclude that, for a two-level system, the optimal process presented here is unique up to global shifts of the energy levels (which we have fixed imposing $E_0^j=0$).
This however is no longer  the case  when operating on $d$-level systems with $d\geq 3$, multiple number of optimal protocols being allowed in this case (see section Methods for details).

\vspace{1 em}
\noindent {\bf  Optimal transformations by one-molecule Szilard-like heat engine}\\
 \begin{figure}[t]
\includegraphics[width=0.44 \textwidth]{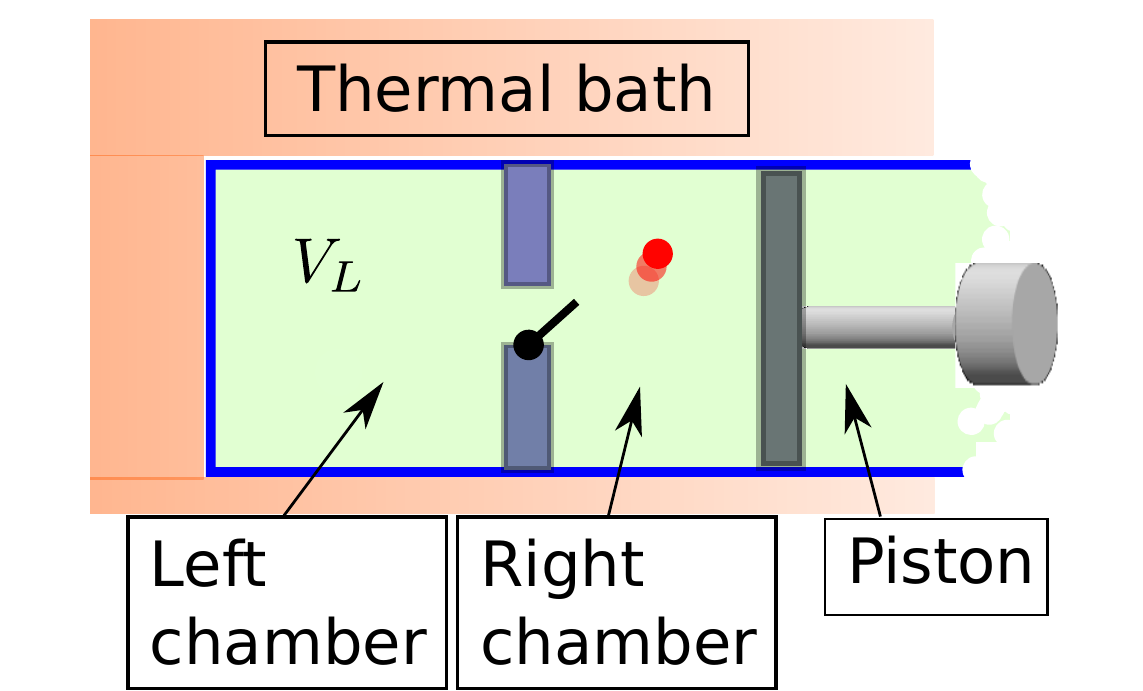}
\caption{An ideal classical system in which we can study the probabilistic extraction of work is composed by a 
box divided in two chambers, with a door regulating the flux of a single particle between them.}\label{fig:piston}
\end{figure}
In this section we present a second example of a process which allows us to saturate the bound~(\ref{bound}). 
At variance with the one introduced in the previous section the model we analyze here is fully classical even though slightly exotic as it assumes 
the existence of an ideal gas composed by a single particle (same trick adopted in Ref.~\cite{SZILARD}). 
As shown in Fig.~\ref{fig:piston} such a classical particle is placed in a box divided in two chambers by a wall, in which a little door can be opened allowing the particle to switch from a side to the other.
The right edge of the box is connected to a piston, that can extract mechanical work, and the whole system is in contact with a heath bath of
temperature $T$.

\begin{figure}[t]
\includegraphics[width=0.42 \textwidth]{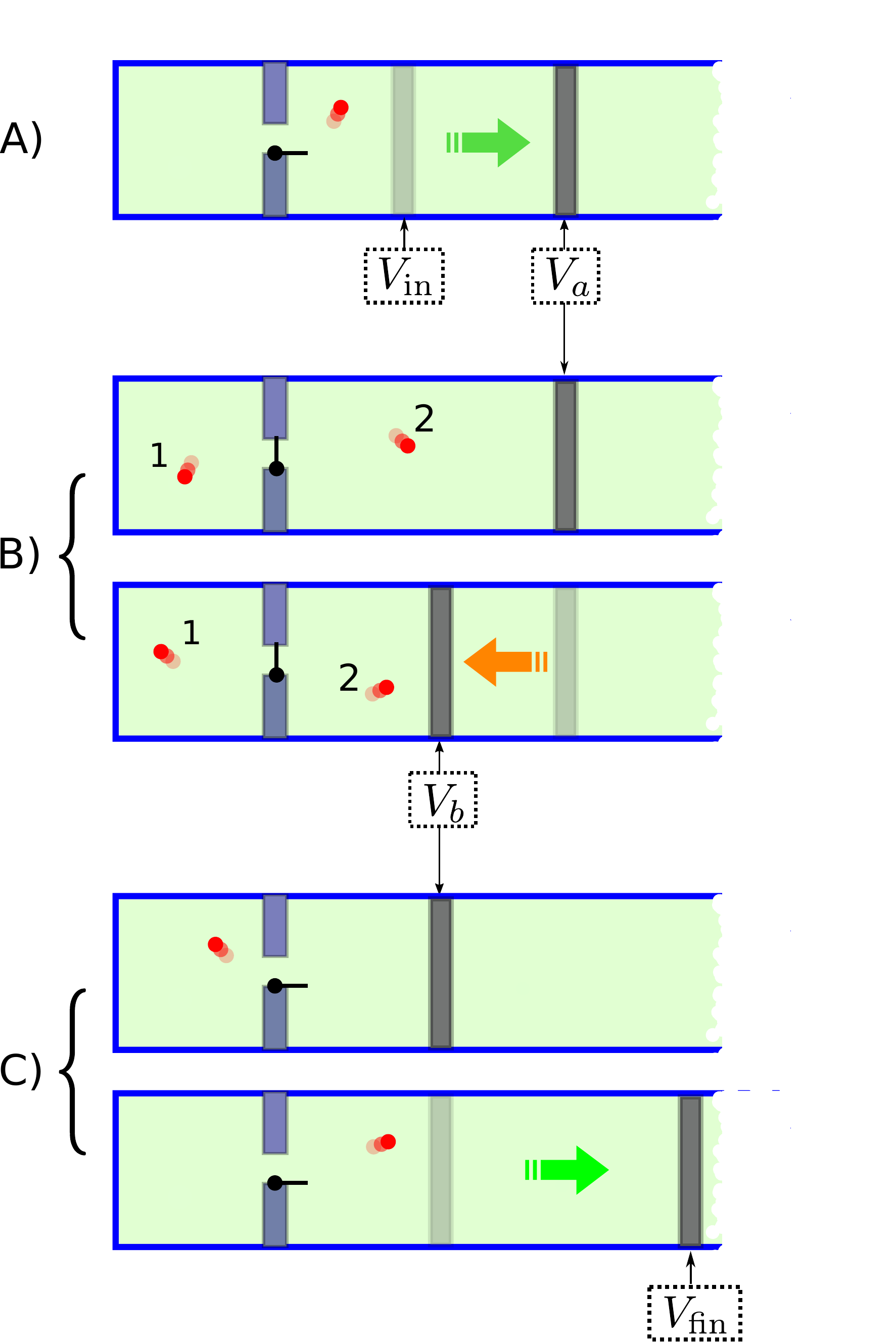}
\caption{Scheme of the optimal process saturating the bound (\ref{bound}) for a one particle perfect gas. The
process is divided in three steps: a) a reversible isothermal expansion from $V_{\rm in}$ to $V_a$ in which the door is \textit{open}, b) a 
reversible isothermal compression between $V_a$ and $V_b$ in which the door is \textit{closed}, c) a reversible isothermal 
expansion to the final volume $V_{\rm fin}$ in which the door is \textit{open}. }\label{fig:gastrasform}
\end{figure}

In the initial and final configuration the door is open, the only difference being the position of the piston.
The free energy difference can be computed following the relation $F= U-TS$ and the fact that the entropy depends logarithmically by the volume~\cite{huang1987}:
\begin{equation}  \Delta F = - \frac{1}{\beta} \log \left(\frac{V_{\rm fin}}{V_{\rm in}} \right). \label{DEFFF}  \end{equation}
We will show that, in order to saturate the equation (\ref{bound}) with the above initial and final conditions the optimal protocol is the following (see Fig. \ref{fig:gastrasform}):

\begin{enumerate}
 \item[a)] keeping the door opened, perform a reversible isothermal expansion from the volume $V_{\rm in}$ to the volume $V_a$ to be fixed later on;
 \item[b)] after closing the door, do a reversible isothermal compression from the volume $V_a$ to the volume $V_b$, then open
 the door and let the system thermalize;
 \item[c)] perform a reversible isothermal expansion from the volume $V_b$ to the volume $V_{\rm fin}$.
\end{enumerate}

The work extracted during the isothermal expansions a) and c) is
\begin{equation} \label{gasmax} W_{a+c} = \frac{1}{\beta} \log \left(\frac{V_a}{V_{\rm in}}\right) + \frac{1}{\beta} \log \left(\frac{V_{\rm fin}}{V_b}\right),\end{equation}
On the contrary to compute the work extracted in the compression we have to distinguish two cases:
\begin{enumerate}
 \item The particle is in the left side, then the compression requires no work.
 \item The particle is in the right side then the work extracted is the negative quantity
  \begin{equation} \label{gasvar} W_b = \frac{1}{\beta} \log \left( \frac{V_b-V_L}{V_a-V_L} \right),\end{equation}
   where $V_L$ is the constant volume of the left chamber.
 \end{enumerate}
 Now observing that the probability of the particle being in the left chamber is just equal to the ratio between 
 $V_L$ and $V_a$, i.e. $p = {V_L}/{V_a}$
the work distribution of the process can be expressed as:
\begin{equation} P(W) = p  \delta(W- W_{\max}) + (1-p ) \delta(W- W_{\min}),   \end{equation}
with
\begin{eqnarray}  \label{DEFWMA} 
W_{\max} := W_{a+c} \;, \qquad W_{\min} := W_{\max} + W_b\;.\end{eqnarray} 
Notice also that from Eqs.~(\ref{DEFFF}),  (\ref{gasmax}), (\ref{gasvar}) and (\ref{DEFWMA}) we can cast $p$ as in Eq.~(\ref{PPPP}), indeed
\begin{equation} p= \frac{V_L}{V_a} = \dfrac{ \dfrac{V_b}{V_a} - \dfrac{V_b - V_L}{V_a - V_L}}{1 - \dfrac{V_b - V_L}{V_a - V_L}} 
= \frac{e^{-\beta \Delta F} - e^{\beta W_{\min}}}{e^{\beta W_{\max}} - e^{\beta W_{\min}}},  \label{calcp}
 \end{equation}
where the last equality is obtained multiplying both the numerator and 
the denominator by $\frac{V_f V_a}{V_i V_b}$.
Then, as in the case of the discrete quantum process analyzed in the previous section, this protocol saturates the inequality~(\ref{bound})
by simply setting $W_{\max}= \Lambda$, the  values of $V_a$ and $V_b$ being univocally fixed by the relations: 
\begin{eqnarray}  V_a &=& V_L \frac{e^{\beta W_{\max}} - e^{\beta W_{\min}}}{   e^{-\beta \Delta F}-e^{\beta W_{\min}}},  \label{Vagas} \\
 V_b &=& V_L  \frac{e^{-\beta W_{\max}} - e^{-\beta W_{\min}}}{e^{\beta \Delta F} - e^{-\beta W_{\min}}},  \label{Vbgas}   \end{eqnarray}
whose positivity is guaranteed once again by the ordering~(\ref{ordering}). 

The protocol presented here clearly shares strong similarities with the two-level model presented in the previous section. 
Indeed the two reversible isothermal expansions in which the particle is free to go throw the door can be put in a formal correspondence with the two quasi-static transformations of Fig.~\ref{fig:gastrasform}. 
Analogously  the intermediate compression of Fig.~(\ref{fig:gastrasform}) corresponds to the  finite DUQ of the quantum 
model.
Notice finally that, since in the ideal gas model the closing of the door at stage b) is a reversible operation,  it looks like we are extracting work in a reversible way 
over the threshold $-\Delta F$, a fact which is impossible~\cite{huang1987}.
This however is not the case since the thermalization that follows the opening of the door after the 
compression makes  the process globally irreversible and the probabilistic outdoing of $-\Delta F$ is fully justified.\\

\vspace{1 em}
\noindent {\bf \large Work extraction  above threshold under average work constraint}\\
%\label{sec:AVE}

\begin{figure}[t]
\includegraphics[width=0.44 \textwidth]{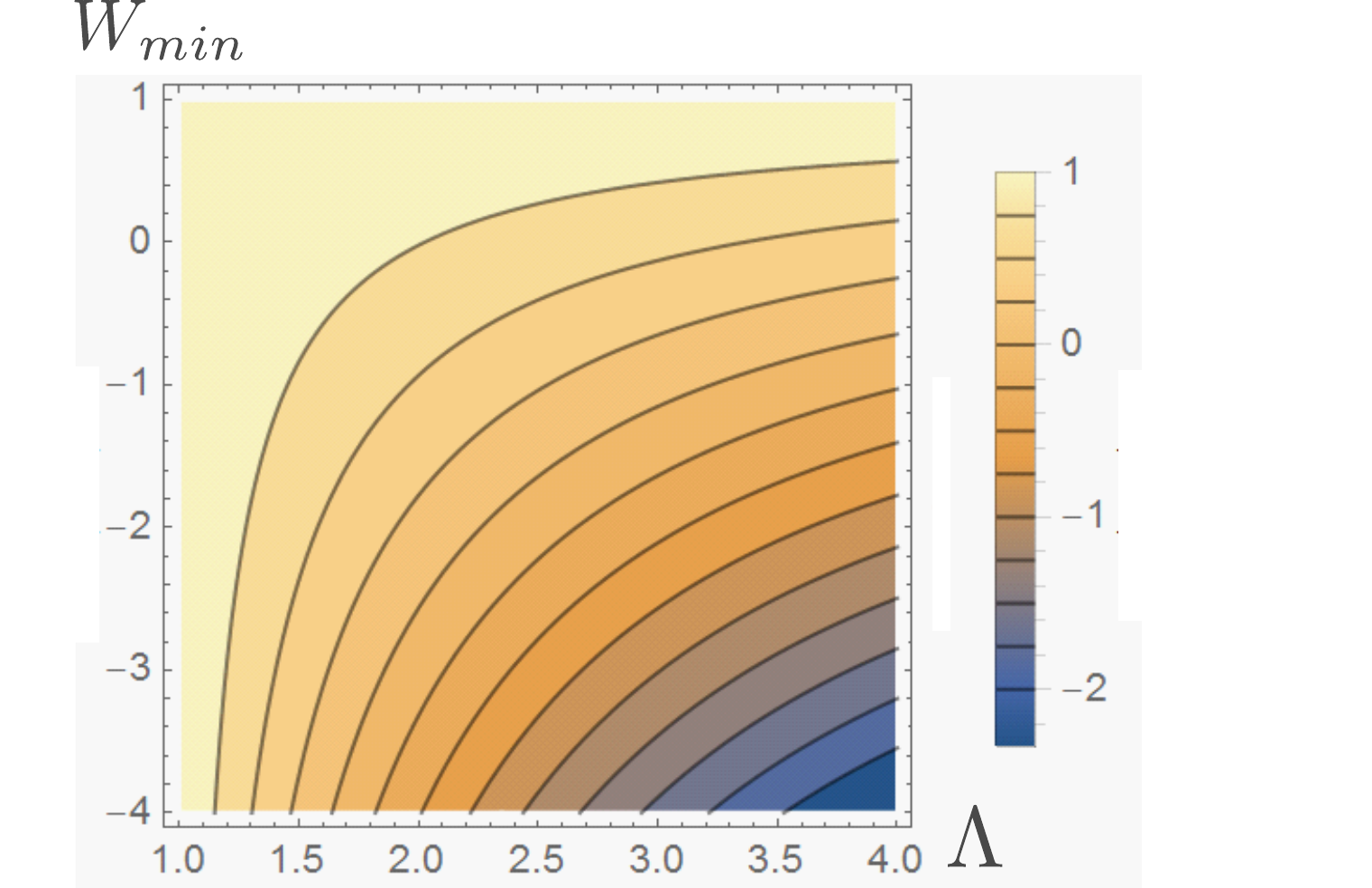}
\caption{Contour plot of the average work $\mu$ for $\beta =\frac{1}{2}$ and 
$-\Delta F =1$, as a function of $W_{\min}$ and $\varLambda$. 
The plot clarifies the increasing behaviour of $\mu$ as a function of $W_{\min}$ for every fixed value of $\Lambda$.}
\label{fig:media}
\end{figure}

\begin{figure}[t]
\includegraphics[width=0.45 \textwidth]{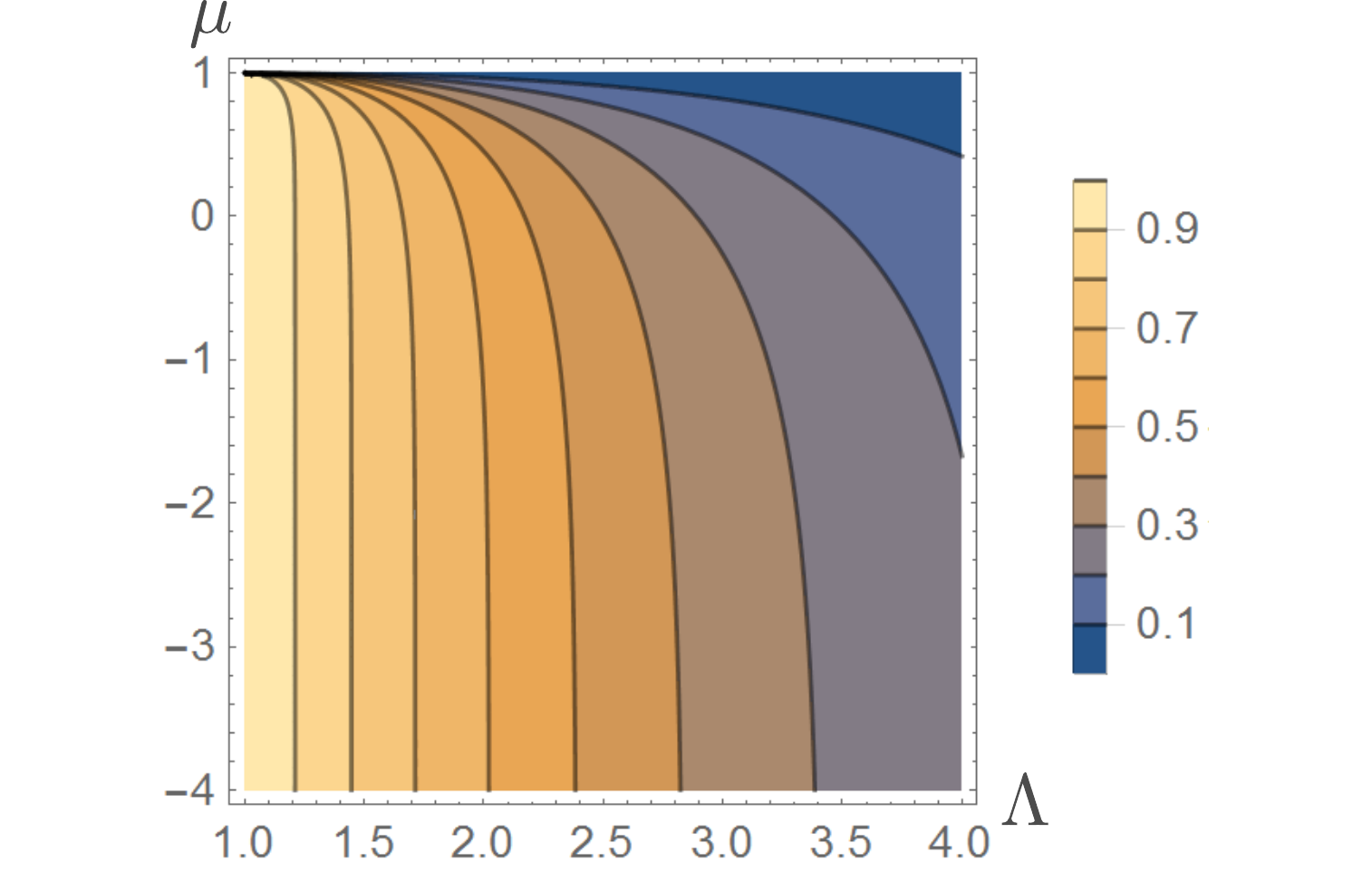}
\caption{Contourplot of the optimal probability $P_{\max}^{(\mu)}(W \ge \varLambda)$ of Eq.~ (\ref{Pmax1111}) for $\beta =\frac{1}{2}$ and 
$-\Delta F =1$, as a function of $\mu$ and $\varLambda$. By definition $\mu$ can not exceed $-\Delta F$ and the problem for $\Lambda \leq -\Delta F$
becomes trivial, so we excluded those regions from the plot.}
\label{fig:contourmedia}
\end{figure}

In the previous section we derived a bound for the probability $P(W>\Lambda)$ when the minimum extracted work $W_{\min}$ is 
fixed.
We are going to solve the same problem with a different constraint, fixing the average extracted work instead of the minimum, i.e. 
replacing Eq.~(\ref{constr1}) with the condition 
\begin{equation} \langle W \rangle = \int_{-\infty}^{\infty} P(W) W dW = \mu,  \label{average}  \end{equation}
with $\mu \leq \Lambda$ being a fixed value.
As we shall see in the following this problem admits optimal processes which have the same dichotomic structure as the optimal solutions
one gets when imposing the constraint on the minimal work production. To be more precise Eqs.~(\ref{PWtot2}) and~(\ref{poptimal})
still provide the optimal solutions by setting $W_{\min}$ to fulfil Eq.~(\ref{average}), i.e. solving for the following transcendental equation 
\begin{eqnarray} \label{mu}  && \mu =  p \; \Lambda + (1- p) \;W_{\min}  =\frac{(e^{-\beta \Delta F} - e^{\beta W_{\min}}) \Lambda + (e^{\beta \Lambda} - e^{-\beta \Delta F} ) W_{\min}}{e^{\beta \Lambda} - e^{\beta W_{\min}}}. \label{mulambda} 
 \end{eqnarray}
Noticing that $\mu$ is strictly increasing in $W_{min}$ for every fixed $\Lambda$ (see Fig. \ref{fig:media}) and so there exists one and only one value $W_{\min}[\Lambda, \mu]$ that solves equation (\ref{mu}).
We can then conclude that the optimal probability in this case is given by the function
\begin{equation}
P_{\max}^{(\mu)}(W \ge \varLambda) \label{Pmax1111} = \frac{ e^{- \beta \Delta F} - e^{\beta W_{\min}[\Lambda,\mu]}}{e^{\beta \varLambda} -e^{\beta W_{\min}[\Lambda,\mu]}},
\end{equation}
which we have plotted in  Fig.~\ref{fig:contourmedia} for fixed values of $\Delta F$ and $\beta$. 

To prove the above results we adopt the Lagrange multiplier technique to study the stationary points of Eq.~(\ref{Pmax}) under the constraint (\ref{average}). Also, to avoid technicalities we find it convenient to discretize the probability distribution, a trick which allows us to 
impose the positivity of $P(W)$ by parametrizing it as $q^2(W)$ with $q(W)$ being an arbitrary function. 
Accordingly 
the  Lagrangian of the problem can be written in this way:
\begin{equation} 
\mathcal{L} =  \sum_{W \geq \Lambda} p(W) + \lambda_{J} \left[\sum_{W} e^{\beta W} q^2(W)  - e^{-\beta \Delta F}\right]   + \lambda_{\mu} \left[ \sum_{W} q^2(W)  W - \mu \right] + \lambda_1 \left[\sum_{W}q^2(W)  -1\right].   
\end{equation}
where $\lambda_{J}$,$\lambda_{\mu}$ and $\lambda_1$ are the Lagrange multipliers that enforce, respectively,  the Jarzynski identity (\ref{jarzynski}), 
 the average constraint~(\ref{average}),
and the normalization. 
Deriving with respect to $q(W)$ we hence obtain the following  Lagrange equation:
\begin{equation} q(W)[ \Theta(W-\Lambda) +\lambda_{J} \beta e^{\beta W} + \lambda_\mu W + \lambda_1] =0, \label{supp} \end{equation}
with $\Theta(W)$ being the Heaviside step function. The above identity must hold for all $W$: 
accordingly the supports of $q(W)$ and of the function in the square brackets must be complementary.  This last term
nullifies in at most three points (say $W_1$, $W_2$, and $W_3$), only one of which (say $W_3$)  is above the threshold $\Lambda$ (verifying this 
property is easy, since the zeroes are 
the crossing points between an exponential function and a piecewise linear function).
Thus we can suppose that $q(W)$, and hence the corresponding probability distribution $P(W) = q^2(W)$,  is non zero on only these selected  points.
Indicating 
with $p_1$, $p_2$ and $p_3$ the values assumed by $P(W)$ on $W_1$, $W_2$ and $W_3$ we  can then express the problem  constraints as follows 
\begin{equation} p_1 + p_2 + p_3 = 1 , \label{prob} \end{equation}
\begin{equation} p_1 W_1 + p_2 W_2 + p_3 W_3 = \mu ,  \label{media3} \end{equation}
\begin{equation} p_1 e^{\beta W_1} + p_2 e^{\beta W_2} + p_3 e^{\beta W_3} = e^{-\beta \Delta F}. \label{jarz3} \end{equation}
Among all possible solutions of these last equations we have finally to select those which provide the maximum value for the probability
of extracting work above the threshold $\Lambda$, i.e. remembering that 
 only $W_3$ can be larger than or equal to $\Lambda$, this corresponds to select the solution with the largest value of $p_3$. 
 To solve this last problem we resort once more to the Lagrange multiplier technique under 
the Karush-Kuhn-Tucker conditions \cite{Boyd2004} to enforce the positivity of $W_3 - \Lambda$. Accordingly the new
Lagrangian is now 
 \begin{equation}   \mathcal{L}'  =  p_3 + \lambda'_J \left[ \sum_{i=1}^3 e^{\beta W_i} p_i - e^{-\beta \Delta F} \right]   +
 \lambda'_{\mu} \left[\sum_{i=1}^3 p_i W_i - \mu \right] + \lambda'_1 \left[\sum_{i=1}^3 p_i -1 \right]+ \eta ( \Lambda- W_3),      
 \end{equation}
 with 
 \begin{equation}  \eta(\varLambda - W_3) =0\quad and\quad \eta \geq 0.  \label{2cond} \end{equation}
The KKT conditions are necessary (even if not sufficient) for a point to be a costrained maximum, and they allow two kind of solutions:
\begin{enumerate}
\item The maximum is in the internal part of the region described by the inequality constraint, then $ W_3 \neq \varLambda$ and by
the conditions (\ref{2cond}) we obtain $\eta =0$.
In this case the Lagrange equations relative to $W_i$ ($i=1,2,3$) are:
\begin{equation} p_i[ \beta \lambda'_{J} e^{\beta W_i} + \lambda'_{\mu}]=0,   \label{lag}  \end{equation}
then either
$W_1=W_2=W_3>\varLambda$ or, if one or more of the $p_i$ are equal to zero, the $W_k$ for $k\neq i$ are all equal.
These solutions has to be rejected, because fails to satisfy the constraint (\ref{media3}).
\item The maximum is on the boundary of the inequality constraint then $W_3 = \varLambda$.
From the conditions (\ref{2cond}) $\eta$ can be different 
from 0, then, altough the Lagrange equations for $W_1$ and $W_2$ are still described by the (\ref{lag}) the one for $W_3$ is not.
Thus either $W_1=W_2 \leq \varLambda = W_3$ or one between $p_1$ and $p_2$ vanishes.
In both cases the support of the work distribution reduces to two points.
\end{enumerate}
We conclude that in order to maximize $P(W\geq \varLambda)$ when the
average extracted work $\mu$ is fixed, the distribution must be different from zero in only two values, $\Lambda$, and a smaller one we call $W_{\min}$, i.e. as anticipated at the beginning of the section, they must have the form (\ref{PWtot2}) yielding Eq.~(\ref{Pmax1111}) as the optimal probability of work extraction above threshold. \\

\vspace{1 em}
\noindent {\bf \large Discussion}\\

In this paper we studied single-shot thermodynamic processes focusing on the specific task of probabilistically extracting more work than what is allowed by the second law of thermodynamics. 
We found that for all processes obeying the Jarzynski identity, there exists an upper bound \eqref{bound} to the work extraction probability which depends on: how large is the desired violation, the minimum work that we are willing to extract in case of failure, and the free energy difference between the final and initial states. Moreover, within the formalism of discrete quantum processes, we have shown that the bound can be saturated and we determined the corresponding optimal protocols. Analogous results have been obtained  also when replacing the constraint on the minimal
work with a constraint on the average work extracted during the process. 

With our analysis we hope to contribute to the yet quite unexplored regime \cite{evans1993,wang2002, jarzynski2011, alhambra2015, halpern2015} in which statistical fluctuations are not considered as a problem but as an advantage of microscopic thermodynamics, in the sense that they can be artificially enhanced in order to obtain tasks which are otherwise impossible in the thermodynamic limit. 
With respect to standard thermodynamics, in this regime we should adopt a completely different paradigm for judging what is a ``good" process. Indeed quasi-static processes are usually considered as optimal since they are reversible, they do not produce excess entropy, they allow to reach the Carnot efficiency, {\it etc.}.  On the other hand, as we have shown in this work, in specific regimes in which fluctuations are ``useful" the perspective is reversed and non-equilibrium processes becomes operationally optimal.

Our findings could be experimentally demonstrated in every classical or quantum thermodynamics experiment involving large work fluctuations. For example experimental scenarios which are currently promising are:  organic molecules \cite{collin2005, liphardt2002}, NMR systems \cite{toyabe2010, batalhao2014}, electronic circuits \cite{saira2012,pekola2015},    trapped ions \cite{an2015}, colloidal particles \cite{berut2012}, {\it etc}. In all these contexts, up to now the task has been mainly focused on the verification of  quantum thermodynamic principles and fluctuation theorems. We believe that, similar experimental settings can be easily optimized in order to maximize the probability of work extraction, approximately realizing the ideal processes proposed in this work.\\

\vspace{1 em}
\noindent {\bf \large Methods}\\

\noindent {\bf  Jarzynski identity for the discrete process}\\
Here we show that  the discrete  process
 described by the sequence~(\ref{steps}) satisfies the Jarzynski identity~\eqref{jarzynski}. Ultimately  this is 
a direct consequence of the fact that each single DUT + DTT process fulfills such relation \cite{tasaki2000}.
Indeed from Eq.~(\ref{PDUQ1}) we can write 
\begin{eqnarray}
&&\!\!\! \!\!\!\!\!\! \langle e^{\beta W} \rangle =   \int d W P(W) e^{ \beta W} = \int d W_1 \cdots \int  dW_N       P_1(W_1) \cdots P_N(W_N) 
 e^{\beta (W_1 + 
\cdots + W_N)}  \nonumber \\
&& \!\!\!= \sum_{k_1, \cdots, k_N}  p_{k_1}^{(0)}\cdots p_{k_N}^{(N-1)}  e^{-\beta(\Delta E_{k_1}^{(1)}  + \cdots  +
\Delta E_{k_N}^{(N)})} = \dfrac{\sum_{k_1, \cdots, k_N}  e^{- \beta (E_{k_1}^{(1)} + \cdots + E_{k_N}^{(N)})}}{Z_0 \cdots Z_{N-1} }\\ \nonumber 
&& \!\!\!= \frac{Z_N}{Z_0}=\exp[{- \beta (F_N - F_0)}] = \exp[ -\beta \Delta F],
\end{eqnarray}
where in the last identities we used Eq.~(\ref{DEFP}) and the fact that the partition function $Z_j$ of the Gibbs state 
$\omega^{(\beta)}_{H_{j}}$ is  connected to its Helmholtz  $F_j$  free energy via the identity $Z_j =\sum_k  e^{- \beta E_{k}^{(j)}}=  e^{{-\beta F_j}} .$

\noindent {\bf  Optimal processes for discrete transformation with  $d>2$ level systems}\\
In the main text we have shown that a two level system is already enough to achieve the maximum work extraction probability dictated by the bound~(\ref{bound}). However, for experimental reasons, one may be forced to work with a system characterized by $d>2$ energy levels, 
and we may be interested in determining what are the optimal processes in this context. 
In the main text we noticed that the presence of only one quench is a necessary condition to saturate the
bound (\ref{bound}). That kind of reasoning holds for systems of arbitrary dimension, from which Eq.~(\ref{PWtot2}) is obviously independent. Thus, looking for an optimal process involving a $d$-dimensional system  we have to slightly modify the procedure described in the main text although
the structure remains the same. Explicitly: 

\begin{enumerate}
 \item Perform a quasi-static transformation that brings all the initial  energies eigenvalues $E_k^{(0)}$ of the system to their final values, except one of them (say the
 $m$-th one), that is instead brought to the value  $E_a$ which will be fixed later on, i.e. 
 \begin{eqnarray} &&E^{(0)}_k \; \rightarrow   E^{(N)}_0 ,  \quad \forall k\neq m  \nonumber \\
   && E^{(0)}_m \rightarrow E_a.  \end{eqnarray}
 \item Apply a finite DUQ which moves the $m$-th level from $E_a$ to a value
$E_b > E_a$ followed by a complete thermalization of the system. 
 \item Perform a quasi-static transformation that brings $E_b$ to the final value $E_m^{(N)}$, in this way the system reaches the final
 configuration.
\end{enumerate}
As in the  two-level case discussed in the main text the distribution of work 
is given by the sum of two delta functions terms~(\ref{PWtot}) the only difference being in the value of the probability $p_0$ which now is given by
\begin{equation}   p =   \frac{ \sum_{k \neq m} e^{-\beta E_k^{(N)}} }{\sum_{k \neq m} e^{-\beta E_k^{(N)}} + e^{-\beta E_a}} , \label{pD}  \end{equation}
while the difference of the free energies associated with the intermediate steps of the protocol which enters Eq.~(\ref{eqLambda}) is  now expressed as 
\begin{eqnarray} \label{FABG}  && F(E_b) - F(E_a) \label{FD}\\&&\qquad = \frac{1}{\beta} 
\log \left[\frac{ \sum_{k \neq m} e^{-\beta E_k^{(N)}} + e^{-\beta E_b}}{\sum_{k \neq m} e^{-\beta E_k^{(N)}}+ e^{-\beta E_a} \nonumber
}\right]. \end{eqnarray}
We can then calculate $E_a$ and $E_b$ from Eqs. (\ref{eqLambda}), (\ref{eqWmin}) and (\ref{FABG}) obtaining:
\begin{equation} \label{eEaG}   e^{-\beta E_a} =\left[ \sum_{k \neq m} e^{-\beta E_k^{(N)}}\right]  \frac{e^{-\beta \Delta F} - e^{\beta W_{\max}}}{e^{\beta W_{\min}} - e^{-\beta \Delta F}},   \end{equation}
\begin{equation} \label{eEbG}   e^{-\beta E_b} = \left[ \sum_{k \neq m} e^{-\beta E_k^{(N)}}\right]  \frac{e^{\beta \Delta F} - e^{-\beta W_{\max}}}{e^{-\beta W_{\min}} - e^{\beta \Delta F}}.   \end{equation}
Then, computing the success probability with the equation (\ref{pD}) we find that $p$ as exactly the same form we have for the $d=2$ case, i.e. 
Eq.~(\ref{PPPP}). 
As a final remark we notice that the protocol saturating the bound is not unique for $d\ge3$, because in this case there are multiple degrees of freedom in the system. For example there are different equivalent choices of the energy  level  $E_m$, moreover other optimal protocols involving multi-level quenches could exist.

\vspace{1 em}
\noindent{\bf  Acknowledgements}\\
The authors are grateful to J. Anders and M. Campisi for useful discussions. This work is partially supported by the EU Collaborative Project TherMiQ (grant agreement 618074) and 
by the ERC Advanced Grant n. 321122 SouLMan.\\

\noindent{\bf  Author contributions}\\ 
VC, AM and VG equally contributed to the derivation and writing of this work.\\

\noindent{\bf  Competing financial interests}\\ 
The authors declare no competing financial interests.

\end{document}